\documentclass[aps,prb,preprint,groupedaddress]{revtex4}

\usepackage[dvips]{graphicx}
\begin{document}

\title{Spin torque switching in perpendicular films at finite temperature, HP-13}


\author{Ru Zhu}
\email{rzhu1@bama.ua.edu}
\author{P. B. Visscher}
\email{visscher@ua.edu}
\affiliation{MINT Center and Department of Physics and Astronomy\\
University of Alabama, Tuscaloosa AL 35487-0209}

\date{\today}

\begin{abstract}
We show how the phase diagram for spin torque switching in the case
of perpendicular anisotropy is altered at nonzero temperature. The
hysteresis region in which the parallel and antiparallel states
coexist shrinks, and a new region of telegraph noise appears. In a small sample, the region of coexistence of a precessional and parallel state can disappear entirely.
We show that the phase diagram for both zero and nonzero temperature
can be understood and calculated by plotting an effective energy as
a function of angle. A combinatorial analysis is useful for systematically describing the phase diagram.
\end{abstract}


\maketitle

\section{Introduction}
In this paper we consider a thin film magnetic element with
perpendicular anisotropy.  The phase diagram for this system has
been studied theoretically \cite{bazaliy} at zero temperature on the
assumption of a homogeneous single domain, and experimentally
\cite{fullerton}. Some discrepancies appear to be due to
inhomogeneities, but some may be due to the fact that the
zero-temperature theory does not take into account thermal
fluctuations.  The latter effects can be calculated
semi-analytically within the single domain model, whereas
inhomogeneity effects probably can only be dealt with numerically.
Thus, in this paper we will generalize the single-domain phase diagram to nonzero temperature; to the best of our knowledge, this has not been done previously.

In uniaxial symmetry, the energy depends only on the angle $\theta$ of the magnetization from the easy axis:
\begin{equation}
E(\theta) = M_s [H_K ^{\textrm{eff}} \sin^2 \theta - H_e \cos \theta ]
  \label{E}
  \end{equation}
where $M_s$ is the saturation magnetization, $H_K ^{\textrm{eff}}=H_K - M_s$ is the effective anisotropy, and $H_e$ is an external field along the easy axis (normal to the film).  Any discussion of spin torque dynamics begins with the Landau-Lifshitz (LL) equation\cite{FP} for the time derivative of the magnetization, $\dot{\mathbf{M}}$, sometimes loosely referred to as "torque".
The LL equation has a precession term which contributes only to the azimuthal component $\dot{\mathbf{M_\phi}}$ and doesn't change the energy; the changes in energy are controlled by the $\theta$ component
\begin{equation}
 \dot M_\theta = %
 M_s \dot{\theta} =
 -\gamma \sin \theta \{ \alpha M_s [H_K ^{\textrm{eff}} \cos \theta + H_e] + \frac{ J M_s}{(1+B \cos \theta)}\}
\label{torque}
\end{equation}
where $\gamma$ is the gyromagnetic ratio.
The first term (proportional to the LL damping parameter $\alpha$) controls damping and pushes $\mathbf{M}$ toward the easy axis, and the second (spin torque) term scales with a parameter $J$ proportional to the current ($B$ is Slonczewski's dimensionless torque asymmetry parameter\cite{slon96})

\section{Effective Energy}\label{EffEn}
Defining an effective energy in the presence of spin torque is nontrivial, since only the precession term in the Landau-Lifshitz equation conserves energy; the damping and spin torque terms are non-conservative.

The most rigorous way to derive an effective energy is by finding a steady state solution of the Fokker-Planck equation\cite{FP,LiZhang}, which has the form $\exp(-E_{\textrm{eff}}/k_B T)$ where
\begin{equation}
E_{\textrm{eff}}= M_s [\frac{1}{2}H_K ^{\textrm{eff}} (1 - \cos ^2 \theta ) - H_e \cos \theta  - \frac{J}{{\alpha B}}\ln (1 + B\cos \theta )] \\
\label{Eeff}
\end{equation}

It is worth noting, however, that Eq. \ref{Eeff} can be obtained heuristically; in this special case of uniaxial symmetry, if we compute the "work" done by the "torque" (Eq. \ref{torque}), the result is proportional to the effective energy: $\int {\dot M_\theta d\theta } = \alpha \gamma E_{\textrm{eff}}$.

We will use a non-dimensional form $e_{\textrm{eff}}=E_{\textrm{eff}}/ \mu_0 H_K^ {\textrm{eff}} M_s$ of the effective energy,
\begin{equation}
e_{\textrm{eff}}= \frac{1}{2}(1 - u^2) - h_e u  - \frac{j}{B}\ln(1+Bu) \\
\label{e}
\end{equation}
where $u=\cos \theta$, $h = H_e/H_K ^{\textrm{eff}}$, and $j=J/\alpha H_K ^{\textrm{eff}}$.

\section{Zero-temperature phase diagram}\label{T0}
\begin{figure}[tbh]
\begin{center}
\includegraphics[width=3.5 in, keepaspectratio,clip]{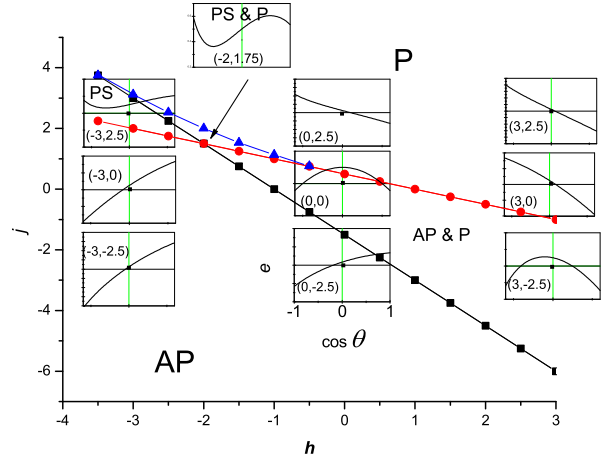}
\end{center}
\par
\vspace{-0.3 in} 
\caption{The phase diagram in the h-j plane for the perpendicular
spin-torque element described in the text, for spin-torque parameter
$B = 0.5$. Lines with symbols are phase boundaries, described in text.  Insets show $e$ \textit{vs.} $u=\cos \theta$
at a $3 \times 3$ grid of points at $h=-3.0, 0, +3.0$ and $j  =
-2.5, 0, +2.5$, plus one additional inset for the small PS \& P region. }\label{mosaic}
\end{figure}

The behavior of Eq. \ref{e} for various values of field and
current is shown in Fig. \ref{mosaic}.  The only free parameter is the Slonczewski parameter $B$, which we have taken to be 0.5 in the figures.  The horizontal and vertical
axes are the dimensionless magnetic field and current, $h$ and $j$, and at each point of a $3 \times 3$ grid there
is an inset graph showing the effective energy as a function of $u =
\cos \theta$, at the field and current corresponding to the center of the inset.
The center graph, at zero field and current, shows only the
anisotropy energy $-u^2/2$, a parabola.  At this
point there are two stable states (minima of $e(u)$ at $u=\pm 1$ -- note that $e(u)$ need
not be flat at minima that lie at the boundaries of the physical
region $-1<u=\cos \theta<1$).  Thus in this region of the phase
diagram both the parallel ($\cos \theta=1$) and antiparallel
($\cos \theta=-1$) states are stable, so we have labeled it "AP \& P".

If we move to the right from the center of Fig. \ref{mosaic}, we add
the Zeeman term $- h u$, which simply shifts the
parabola to the left so its maximum is outside the physical region,
and there is only one (parallel) minimum at $u=1$, and we have
crossed a phase boundary (the red circles -- colors online) to the "P" region where only
the parallel state is stable.  At this boundary, the maximum is
just leaving the physical region at $u=-1$, \textit{i.e.}, $de(u=-1)/du=0$.
On the left, at $h=-3.0$, the parabola shifts in the opposite
direction (right) -- the maximum passes out of the physical region
at the black (square symbols) phase boundary, where $de(u=1)/du=0$,
and only the AP state is stable.

Moving up from the center of Fig. \ref{mosaic} (increasing the
scaled current to $j = 2.5$), the effective-energy inset graph
includes the logarithmic spin-torque term $ - \ln(1+Bu)$.  This has a
divergence at $u = -1/B = -2$, outside the physical region, but
starts to rise at the left as seen in the top center inset.  This
mimics the effect of a positive field (right inset) and brings
us across the phase boundary to the upper-right single-phase P region.

At the upper left inset (negative field $h=-3.0$), the effect
of the field (lowering $e$ at the left) and the spin
torque (raising $e$ at the left) oppose each other,
and the spin torque can raise the energy at the left enough to
create a minimum away from the $u=\pm 1$ boundaries, physically
corresponding to a precessional state -- thus this region of the
phase diagram is labeled "PS".  The effective energy in the
remaining small sliver of the phase diagram, between the curved line (blue triangles)
and the straight phase boundaries, is shown as a tenth inset -- both the precessional and parallel states are
stable, so this region is labeled "PS \& P".

Making the current negative (bottom center inset) causes a decrease in energy at the left
side of the inset, mimicking a negative field.  The only minimum is now the AP (left) minimum, and the system is below the black (squares) phase boundary in the AP region.
At the lower left (adding a real negative field) further stabilizes the AP state.  Moving to the lower right (positive field) restores the $u=1$ (parallel) minimum and returns us above the black phase boundary to the AP \& P region

\section{Systematic combinatorial construction of phase diagram}
\begin{figure}[]
\begin{center}
\includegraphics[width=3.5 in, keepaspectratio,clip]{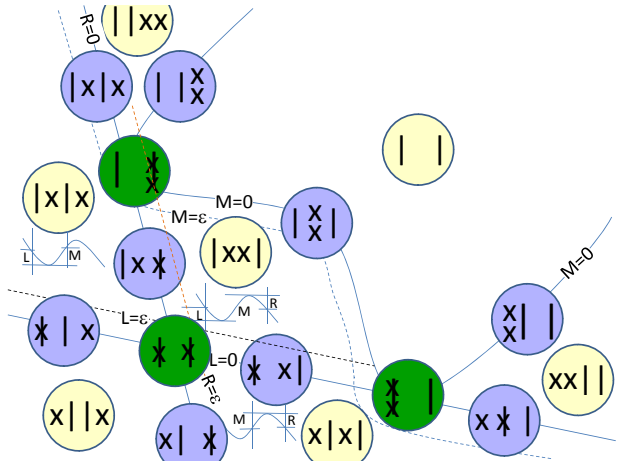}
\end{center}
\par
\vspace{-0.3 in} 
\caption{Cartoon version of the phase diagram (Fig. \ref{mosaic}).  The straight phase boundaries are represented by solid straight lines, but the parabola is distorted into a curved W shape to give room between it and the lines for the symbolic labels.  Light grey (yellow online) circles label regions, darker (blue online) circles label boundaries, and darkest (green online) circles label intersection points. $\times$'s are extrema of $e(u)$, $|$ and $|$ are $u=-1$ and $1$; one $\times$ above the other means the minimum and maximum coincide (this occurs on the parabola $M=0$). In the three regions in which there are minima separated by barriers, a sketch of $E$ \textit{vs} $\cos \theta$ is shown, to define the left, middle, and right (L, M, and R) barriers.}
\label{cartoon}
\end{figure}

It is not obvious from the above approach to the phase diagram how many regions there can be.  We now enumerate the regions more systematically.  The physical behavior depends on the number and position of extrema of the effective energy $e(u)$.  The condition $de/du=0$ gives a quadratic equation for u, with two solutions $u^-<u^+$ if $j<\frac{B}{4}(h-\frac{1}{B})^2$, i.e. below the parabola shown by blue triangles in Fig. \ref{mosaic}.  Above this parabola there are no flat maxima or minima (as opposed to the boundary extrema at $u=\pm 1$), and the boundary minimum always occurs at u=1: this region is a parallel (P) state.  The lowest straight line [black squares, $j=-(h+1)/(1+B)$] in Fig. \ref{mosaic} is where $de(u=1)/du=0$, i.e. one of the extrema occurs at the right boundary, and the other straight line [circles (red online) $j=-(h-1)/(1-B)$] has $de(u=-1)/du=0$.  Below the parabola we classify the positions of the extrema relative to the boundaries $u=\pm 1$ of the physical region by a four-character string such as $|\times \times|$, where the vertical lines represent $u=-1$ and $u=+1$ respectively and the $\times$'s represent the minimum and the maximum (in that order).  There are exactly ($\begin{array}{c}
                    4 \\
                    2
                  \end{array}) \equiv \frac{4!}{2!2!}=6$
ways to pick 2 of the 4 positions for $\times$, i.e. to order these 4 characters; if we add the string $||$ to represent the P region without extrema, these are exactly the 7 light grey (yellow online) circles in Fig. \ref{cartoon}, a distorted cartoon of the phase diagram.
The boundary between each adjoining pair of regions (light circles) is labeled by a darker grey circle (blue online) with a string in which two of the symbols are superposed (e.g., $* \times |$, where $*$ indicates a $|$ and $\times$ superposed, near the bottom center of Fig. \ref{cartoon}, means there is an extremum at the left boundary $u=-1$; there the $|$ and $\times$ pass each other, converting $|\times\times|$ into $\times|\times|$).  Each boundary similarly involves the crossing of two symbols, except that $|$ and $|$ cannot cross, and $\times$ and $\times$ "cross" only when they merge -- beyond this boundary (the distorted parabola) there is no minimum or maximum.  At the intersections of boundaries (darkest grey circles, green online) there are \emph{two} coincident symbols: at the rightmost intersection, $\times$ crosses $*$ -- at that point the minimum and maximum annihilate at $u=-1$.

\section{Phase diagram at nonzero temperature}
At nonzero temperature, the system will not remain in a well with a very low barrier.  If we assume an Arrhenius-Neel model for the switching rate with a prefactor $\nu$, an  experimental time scale $\tau$, and an experimental temperature $T$, the critical value (call it $\epsilon$) of the dimensionless barrier $\Delta e$, below which a well is not stable, is given by %
$\tau \nu \exp(- \mu_0 H_K ^{\textrm{eff}} M_s V \epsilon /k_B T)=1$; here we use the value $0.03$ for $\epsilon$.
Only 3 of the regions have barriers that can trap the magnetization: $|\times|\times$, $|\times\times|$, and $\times|\times|$.  The energy $e(u)$ is sketched in Fig. \ref{cartoon} in each of these regions, showing the barriers at the left (L), middle (M), and right (R).
The zero-T phase boundaries (solid lines) are where $L=0$, $M=0$, and $R=0$.  The dashed lines where $L=\epsilon$, etc., are the $T>0$ phase boundaries.  For example, in the center of Fig. \ref{cartoon} between the $M=0$ and $M=\epsilon$ labels, the parallel and precessional states are both stable at zero temperature, but at $T$ the precessional state will jump the M barrier and only the parallel state is stable -- thus this point is effectively in the P region, and the boundary moves down to the dashed curve.

\begin{figure}[tbh]
\begin{center}
\includegraphics[width=3.5 in, keepaspectratio,clip]{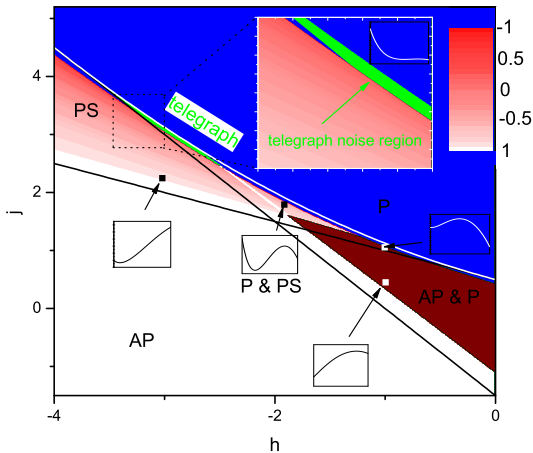}
\end{center}
\par
\vspace{-0.6 in} 
\caption{Nonzero-temperature phase diagram (colored boundaries) -- zero-temperature boundaries are indicated by solid black lines.
Color scale from white to grey (red online) in the PS (precessional state) region and the coexisting parallel and precessional (P \& PS) region indicates $\cos \theta$ for stable precession.
We have colored part of the P region grey (green online) to indicate that it is a physically interesting region -- both the $M$ and $R$ barriers (defined in Fig. \ref{cartoon} are lower than the critical value $\epsilon$, so the system jumps between the precessional and the parallel state rapidly on the experimental time scale -- this is telegraph noise.  Inset at top shows detail of part of this telegraph noise region.  Graphical insets show $e(u)$ energy landscapes at the indicated points, as in Fig. \ref{mosaic}.
}
\label{colors}
\end{figure}

The actual $T>0$ phase diagram is shown in Fig. \ref{colors}.   It is topologically equivalent to the cartoon, but because the boundary shifts are very small in places, the topology is easier to see in the cartoon.  The main result is that the P \& PS coexistence region shrinks, and a region of telegraph noise appears near its boundary.
\vspace{-0.3 in}

\begin{acknowledgments}
\vspace{-0.2 in}
This work was supported by NSF MRSEC grant DMR-0213985 and by the
DOE Computational Materials Science Network.
\end{acknowledgments}


\end{document}